\documentclass[11pt]{article}

\usepackage[preprint]{acl}

\usepackage{times}
\usepackage{latexsym}
\usepackage{hyperref}
 \usepackage{tcolorbox}

\usepackage[T1]{fontenc}
\usepackage[utf8]{inputenc}
\usepackage{xspace}

\usepackage{microtype}

\usepackage{inconsolata}
\usepackage{makecell}

\usepackage{graphicx}
\usepackage{listings}
\usepackage{pifont}

\usepackage{cleveref}
\usepackage{xcolor}

\usepackage{pgfplots}
\pgfplotsset{compat=1.18}

\newcommand{\hlt}[1]{#1}

\newcommand{\name}{\texttt{ExeBench-FHD}\xspace}

\lstdefinestyle{mystyle}{
    backgroundcolor=\color{backcolour},   
    commentstyle=\color{codegreen},
    keywordstyle=\color{magenta},
    numberstyle=\tiny\color{codegray},
    stringstyle=\color{codepurple},
    basicstyle=\ttfamily\footnotesize,
    breakatwhitespace=false,         
    breaklines=true,                 
    captionpos=b,                    
    keepspaces=true,                 
    numbers=left,                    
    numbersep=5pt,                  
    showspaces=false,                
    showstringspaces=false,
    showtabs=false,                  
    tabsize=2,
    xleftmargin=1.2em,
}
\lstdefinelanguage[x64]{Assembler}[x86masm]{Assembler}{
  morekeywords={rax,rbx,rcx,rdx,rsi,rdi,rsp,rbp,r8,r9,r10,r11,r12,r13,r14,r15,xmm0,rip, pxor, cvtsi2ss, addss}
}

\definecolor{codegreen}{rgb}{0,0.6,0}
\definecolor{codegray}{rgb}{0.5,0.5,0.5}
\definecolor{codepurple}{rgb}{0.58,0,0.82}
\definecolor{backcolour}{rgb}{0.95,0.95,0.92}

\lstdefinelanguage{json}{
    basicstyle=\ttfamily\footnotesize,
    numbers=left,
    numberstyle=\tiny\color{codegray},
    stepnumber=1,
    numbersep=5pt,
    showstringspaces=false,
    showspaces=false,
    showtabs=false,
    xleftmargin=1.2em,
    breakatwhitespace=false,
    breaklines=true,
    keepspaces=true,
    tabsize=2,
    backgroundcolor=\color{backcolour},
    stringstyle=\color{codepurple},
    keywordstyle=\color{magenta},
    literate=
     *{0}{{{\color{codegray}0}}}{1}
      {1}{{{\color{codegray}1}}}{1}
      {2}{{{\color{codegray}2}}}{1}
      {3}{{{\color{codegray}3}}}{1}
      {4}{{{\color{codegray}4}}}{1}
      {5}{{{\color{codegray}5}}}{1}
      {6}{{{\color{codegray}6}}}{1}
      {7}{{{\color{codegray}7}}}{1}
      {8}{{{\color{codegray}8}}}{1}
      {9}{{{\color{codegray}9}}}{1}
      {:}{{{\color{codegray}{:}}}}{1}
      {,}{{{\color{codegray}{,}}}}{1}
      {\{}{{{\color{codegray}{\{}}}}{1}
      {\}}{{{\color{codegray}{\}}}}}{1}
      {[}{{{\color{codegray}{[}}}}{1}
      {]}{{{\color{codegray}{]}}}}{1}
      {true}{{{\color{magenta}true}}}{4}
      {false}{{{\color{magenta}false}}}{5}
      {null}{{{\color{magenta}null}}}{4}
}

\title{ExeBench-FHD: Addressing Fidelity Issues in Decompiler Evaluation}

\author{
 \textbf{Akarsh Ashok\textsuperscript{1}},
 \textbf{Dinil Mon Divakaran\textsuperscript{2}},
 \textbf{Vivek Balachandran\textsuperscript{1}},
\\
\\
 \textsuperscript{1}Singapore Institute of Technology \\
 \textsuperscript{2}A*STAR Institute of Advanced Intelligence and Computing
\\
 \small{
   \texttt{2403937@sit.singaporetech.edu.sg},
   \texttt{dinil\_divakaran@a-star.edu.sg},
   \texttt{vivek.b@singaporetech.edu.sg}
 }
}

\begin{document}
\maketitle
\begin{abstract}

ExeBench is one of the most widely used benchmarks for evaluating LLM-based decompilers. We analyze ExeBench and identify nine issues that affect how faithfully a test measures a decompiler. Three of them prevent LLM decompilers from recovering C code from ExeBench's function. These are ambiguous parameter order when integer and floating-point types are mixed, unused parameters that optimized assembly may drop, and constants stored outside the function's assembly code. Six more affect test validity: non-void functions without a return, placeholder types, static or inline functions removed by the optimizer, functions named \texttt{main}, trivial bodies and truncated definitions. We present \name, a backward-compatible revision of ExeBench that adds 16 new fields: call-site assembly, data sections, test wrappers, linked-binary disassembly and five warning flags. In ExeBench, 38.9\% of functions have at least one of the five flags; dropping records with any of the four warning flags (all except \texttt{unused\_param}, which marks functions that the new wrappers make evaluable) keeps 63.9\% of the corpus. In an evaluation at \texttt{-O2} on four architectures, the new fields raise GPT 6 Sol's re-execution rate by up to 26.5\%, 19.1\% and 21.0\% on functions that read data sections, mix parameter types and have unused parameters, respectively.
\end{abstract}

\section{Introduction}

Binary decompilation is the process of recovering source code from a compiled binary, and it is a core step in reverse engineering. Compilation is a lossy process. Information that helps humans understand code, such as variable names, data types, comments and macro definitions, is transformed or removed; the higher the level of binary optimization, the harder the binary is to understand. Commercial and open-source decompilers such as IDA Pro~\citep{hexrays_ida} and Ghidra~\citep{nsa_ghidra} use algorithms and heuristics to recover source code from the binary, but their output still deviates from the original source code.

Large Language Models (LLMs) have recently grown popular not only for coding but also for decompilation. Researchers have proposed approaches that use LLMs in decompilation pipelines \citep{llm_decomp_survey, nova, slade, llm4decompile}. However, only a few datasets contain both source code and the corresponding assembly. These include AnghaBench~\citep{anghabench}, ExeBench~\citep{ExeBench}, HumanEval-Decompile~\citep{llm4decompile}, and, most recently, DecompileBench~\citep{decompilebench} and Decompile-Bench~\citep{decompile-bench}. 

\hlt{Among these, ExeBench is one of the most widely used for training and testing. It contains 6.56M C functions from GitHub and AnghaBench, split into ten training, validation and test sets (\Cref{tab:splits}), and most are stand-alone compilable. Each function comes with assembly for x86 and ARM and with input-output (IO) examples. ExeBench also provides a script for evaluating decompilers on its test splits: the decompiled function is recompiled, linked into a test wrapper that feeds it each example input, and checked against the corresponding expected output.}

%If it does for every example, the test is passed. 
To utilize this test to fairly measure a decompiler, the input should contain what the function needs to be re-created, namely the function's signature (parameter order and count) and any constants it reads. 
%Assembly by itself often does not. 
The wrapper calls the function with its original signature, but ExeBench's assembly of the compiled function may not retain the parameter order, unused parameters and constant values.
This is a problem with the evaluation, not with the decompiler. 

A decompiler can translate the assembly correctly and still fail the test. Functions often require information that a binary in practice has, but the assembly in the dataset does not: \hlt{real-world binaries also contain callers and data sections, whereas ExeBench provides only the function's own body.}  
%When we evaluated decompilers on ExeBench, we found such functions, 
Consequently, a test failure on ExeBench  does not (necessarily) indicate a decompiler made a mistake. We call these \emph{fidelity issues} affecting evaluations.
%. ExeBenchPlus addresses them, in part by providing the missing information together with the assembly.}

We identify nine fidelity issues (\Cref{tab:issue-prevalence}). In three of them, \hlt{the function's assembly in ExeBench lacks information that in practice a  binary provides:} mixed integer and floating-point parameters (P1, 1.5\% of the corpus but 11.5\% of \texttt{test\_real}), unused parameters that the optimizer may remove (P2, 7.6\% of the corpus) and constants stored in data sections (P3, 16.2\% at \texttt{-O2}). Two make the test oracle unreliable: non-void functions without a return (P4, 1.7\%) and placeholder types (P5, 9.3\%). Three prevent a usable source--assembly pair: \texttt{static} or \texttt{inline} functions, which the optimizer may remove at higher optimization levels (P6, 36.3\%); functions named \texttt{main}, which fail to link with the driver \emph{main} (P7, 6.3\%); and truncated definitions, which fail to compile (P9, 0.6\%). Finally, trivial bodies (P8, 27.8\%) are easy to decompile and raise average scores.

To address these issues, we introduce \name\footnote{FHD abbreviates Fidélité de Haute Décompilation, French for "high decompilation fidelity".}, a backward-compatible revision of ExeBench that adds 16 new fields and modifies none of the existing ones. The new fields supply \hlt{the parts of the binary that the function's assembly leaves out} alongside the assembly (\Cref{sec:exebench_audit}): data sections for constants~(P3), call-site assembly (\hlt{the assembly code of a function call}) for parameter order~(P1), and test wrappers that accept either all parameters or only the used ones~(P2). They also provide disassembly regenerated from the linked binary and five warning flags for the remaining issues. Each field that depends on compilation has a diagnostic field recording why it is empty. Because no inherited field changes, existing ExeBench pipelines keep working and can adopt the new fields one at a time (\Cref{sec:exebenchplus_dataset}).

This work makes three contributions:

\begin{enumerate}

    \item \textbf{An analysis of ExeBench.} We analyse all ExeBench functions with automated checks and identify nine fidelity issues. \hlt{In three of them, the function's assembly in ExeBench lacks information that a real-world binary provides (parameter order, unused parameters and constants), which can affect downstream uses such as LLM-based decompilation (\Cref{sec:exebench_audit}).}

    \item \textbf{\name.} \hlt{We release a backward-compatible revision of ExeBench with 16 new fields that supply the missing context (data sections, call-site assembly, either-form test wrappers), regenerated disassembly and five warning flags (\Cref{sec:exebenchplus_dataset,sec:app-schema}).}

    \item \textbf{An evaluation study.} On test split of ExeBench across four architectures (x86, x86-64, ARM32, ARM64), the new fields improve GPT~6~Sol's re-execution rate by up to 26.5\%, 19.1\% and 21.0\% on functions that read data sections, mix parameter types and have unused parameters, respectively (\Cref{sec:downstream-main}).

\end{enumerate}

\section{Related Work}
\label{sec:related_work}

For the decompilation task, there are only a handful of datasets for both training and testing. \textbf{AnghaBench}~\citep{anghabench} is a collection of 1 million C functions scraped from GitHub. It uses the type-inference tool Psyche-C~\citep{psyche-c} to add missing external dependencies, so that each function can be compiled into an object file. However, since Psyche-C only declares the dependencies, GCC treats them as forward declarations: it produces an object file but cannot link it into an executable.

\citet{ExeBench} observe the need for an executable dataset for machine learning and introduce \textbf{ExeBench}, whose source functions are scraped from GitHub and which extends AnghaBench by adding dependencies to the C code. ExeBench walks through each function's repository to find the relevant dependencies and turns the function into a stand-alone compilable file. If the dependencies are not present in the repository, ExeBench searches a larger GitHub dataset, and if more than one repository has the same definition, it chooses the one with more stars. If the real dependencies are not available on GitHub, synthetic definitions are generated and injected into the code. ExeBench also provides IO examples, produced by an automatic generator that synthesizes inputs for each parameter type from the function signature using a seed and records the outputs. 

\hlt{ExeBench has ten splits: six for training and four for validation and testing (\Cref{tab:splits}). \emph{Real} splits use dependencies found on GitHub, \emph{synth} splits use synthesised ones, and \emph{simple}/\emph{rich} IO separates trivially satisfiable IO pairs from richer ones.}

\begin{table}[ht]
  \centering
  \small
  \begin{tabular}{lr}
    \hline
    \textbf{Split} & \textbf{Functions} \\
    \hline
    \texttt{train\_not\_compilable} & 715,700 \\
    \texttt{train\_real\_compilable} & 885,074 \\
    \texttt{train\_synth\_compilable} & 4,159,373 \\
    \texttt{train\_real\_simple\_io} & 42,769 \\
    \texttt{train\_synth\_simple\_io} & 640,116 \\
    \texttt{train\_synth\_rich\_io} & 107,250 \\
    \texttt{valid\_real} & 2,133 \\
    \texttt{valid\_synth} & 5,000 \\
    \texttt{test\_real} & 2,134 \\
    \texttt{test\_synth} & 5,000 \\
    \hline
    Whole corpus & 6,564,549 \\
    \hline
  \end{tabular}
  \caption{\hlt{The ten ExeBench splits and their number of functions.}}
  \label{tab:splits}
\end{table}

Meanwhile, the authors of LLM4Decompile take another approach: they translate the 164 handwritten Python problems of HumanEval~\citep{eval_llm} and their assertions into C, creating \textbf{HumanEval-Decompile}~\citep{llm4decompile}. 

In 2025, two benchmarks with similar names appeared. \textbf{Decompile-Bench}~\citep{decompile-bench} contains two million binary-source function pairs mined from binaries compiled from permissively licensed GitHub projects. It uses a \emph{source-trace} algorithm based on DWARF~\citep{dwarfstd} and tree-sitter~\citep{tree-sitter} to match each assembly function to its source function. The dataset has an uneven number of assembly versions per source function (1 to 6), and the released data lacks the optimization-level field and generation code, so we cannot determine the cause.

\textbf{DecompileBench}~\citep{decompilebench} selects around 20k functions from 130 OSS-Fuzz~\citep{oss-fuzz} projects, keeping those that a fuzzer can trigger. Like ExeBench, it extracts each function's dependencies and builds a shared library containing it. A decompiled function is rebuilt into a library and counts as correct if re-running the fuzzer with the same seed gives the same output as the original. However, its assembly is generated from object code rather than a linked binary, so call targets and data references remain unresolved relocations. \name avoids this by disassembling linked executables.

\begin{tcolorbox}[colback=gray!8, colframe=black!60, boxrule=0.5pt, left=3pt, right=3pt, top=2pt, bottom=2pt]
Our contribution \name is a high-fidelity revision of ExeBench for evaluating decompilers on optimized binaries: it supplies the information that ExeBench's assembly lacks, adds flags for unreliable test cases, and leaves the original fields unchanged. 
%It introduces neither new source functions nor a new evaluation metric, so it complements, rather than competes with, DecompileBench, Decompile-Bench and HumanEval-Decompile.
\end{tcolorbox}

\section{Fidelity Issues in ExeBench}
\label{sec:exebench_audit}
ExeBench contains both source code and assembly code, and is therefore used for training and testing machine learning models. \hlt{Each record holds 23 fields (\Cref{sec:app-schema}); about 55\% of records cannot be linked into an executable because ExeBench does not capture their dependencies (\Cref{tab:app-compile-link}).} In this section, we analyze the full dataset covering all ten official splits, and report the issues that are crucial for evaluating decompilation. Each issue is numbered, and \Cref{sec:exebenchplus_dataset} addresses it under the same number. \Cref{tab:issue-prevalence} gives the share of the corpus and of \texttt{test\_real} affected by each issue.

\begin{table}[ht]
  \centering
  \small
  \resizebox{\columnwidth}{!}{%
  \begin{tabular}{llrr}
    \hline
    \textbf{ID} & \textbf{Issue} & \textbf{Corpus \%} & \textbf{test\_real \%} \\
    \hline
    P1 & Mixed int/float parameters & 1.46 & 11.48 \\
    P2 & Unused parameter & 7.56 & 7.40 \\
    P3 & Reads a data section (\texttt{-O2}) & 16.18 & 16.31 \\
    P4 & Non-void without return & 1.73 & 0.19 \\
    P5 & Dummy UDT dependency & 9.27 & 0.52 \\
    P6 & \texttt{static}/\texttt{inline} definition & 36.31 & 28.73 \\
    P7 & Function named \texttt{main} & 6.30 & 0.00 \\
    P8 & Trivial body & 27.84 & 27.41 \\
    P9 & Truncated definition & 0.63 & 0.00 \\
    \hline
  \end{tabular}}
  \caption{Prevalence of the nine issues over all 6,564,549 functions and over \texttt{test\_real} (2,134). P3: non-empty data section at \texttt{-O2} on at least one architecture. P6: \texttt{static} or \texttt{inline} modifier.}
  \label{tab:issue-prevalence}
\end{table}

\subsection*{P1: Parameter order not visible in the assembly}
On 32-bit Intel (x86), function arguments are passed on the stack, which preserves their order regardless of type. On 64-bit Intel, 32-bit ARM and 64-bit ARM, however, parameters are passed in different sets of registers depending on their data type. \Cref{tab:archs} shows how each architecture passes parameters~\citep{sysvabi64,aapcs32,aapcs64}. In ExeBench's \texttt{test\_real} split, 11.5\% of functions have mixed parameter types (1.5\% of the corpus). If their parameter order is not recovered correctly, the corresponding test fail. 

\begin{table}
  \centering
  \small
  \begin{tabular}{lcc}
    \hline
    \textbf{Arch.}  & \textbf{int} & \textbf{float} \\
    \hline
    \textbf{x86}  & stack & stack \\
    \textbf{x86-64}  & rdi,rsi,rdx,rcx,r8,r9 & xmm0 - xmm7 \\
    \textbf{ARM32}  & r0 - r3 & s0 - s15 \\
    \textbf{ARM64}  & x0 - x7 & s0 - s7 \\
    \hline
  \end{tabular}
  \caption{Registers for \lstinline|int| and \lstinline|float| arguments (ARM32: hard-float ABI).}
  \label{tab:archs}
\end{table}

If all parameters have the same data type (first two functions of \Cref{lst:intnfloat}), LLM-based decompilers can potentially identify the parameter order from the register sequence. However, if a function mixes data types, such as \lstinline|int| and \lstinline|float| (last two functions of \Cref{lst:intnfloat}), the order cannot be recovered from the function's assembly alone (as in ExeBench). This limitation occurs because each data type uses a separate register sequence. For example, the System V x86-64 Application Binary Interface (ABI)~\citep{sysvabi64} independently allocates integer arguments are assigned to~[\texttt{rdi}, \texttt{rsi}, \ldots{}] and floating-point arguments to~[\texttt{xmm0}, \texttt{xmm1}, \ldots{}].

\begin{lstlisting}[language=C, caption={Functions with same-type and mixed-type parameters. The last two have opposite parameter orders but use the same registers.}, label={lst:intnfloat}]
int   sum(int a, int b);     
// a -> rdi,  b -> rsi
float sum(float a, float b); 
// a -> xmm0, b -> xmm1
float sum(int a, float b);   
// a -> rdi,  b -> xmm0
float sum(float a, int b);   
// a -> xmm0, b -> rdi
\end{lstlisting}

A function taking an \lstinline|int| followed by a \lstinline|float|, and one taking a \lstinline|float| followed by an \lstinline|int|, therefore read the same registers and compile to identical assembly. This comes from the design of the calling convention, and is not a weakness of a particular decompiler or a bug in ExeBench. Consequently, even a perfect decompiler can only guess the parameter order from ExeBench's assembly. This creates an evaluation problem: ExeBench's test calls the function with the original order, but the dataset's assembly provides only the isolated assembly of the function. Real-world binaries include calls that reveal the order, but ExeBench does not provide them. As a result, a correct decompilation with swapped parameters  fails the test.

\subsection*{P2: Unused parameters may be removed from optimized assembly}
\hlt{Some ExeBench functions do not use all of their parameters. In \Cref{lst:unused_param}, \texttt{sum} is declared with three parameters, \texttt{a}, \texttt{b} and \texttt{c}, but the C source only uses \texttt{a} and \texttt{c}. At \texttt{O0}, all three parameters are visible in the function's assembly. At higher optimization levels the callee never reads the unused argument, and the compiler handles it in one of two ways. In some builds it keeps each used parameter in its original register, so the assembly still implies all three parameters (\texttt{a} in \texttt{rdi}, \texttt{c} in \texttt{rdx}). In others it moves the used parameters into the first registers, so the assembly implies only two (\texttt{a} in \texttt{rdi}, \texttt{c} in \texttt{rsi}). Which happens depends on the build, even for the same source function, so a decompiler faithful to the assembly can legitimately recover either form, and a test that evaluates one marks the other as wrong. In a real-world binary, the callers of an externally visible function still pass all three arguments, but ExeBench does not provide them.}

\begin{lstlisting}[language=C, caption={Function with an unused parameter}, label={lst:unused_param}]
static int sum(int a, int b, int c) {
    return a + c; // b is never read
}

// O0: a, b and c are all visible in the assembly
// (edi = a, esi = b, edx = c)

// O1 and above: one of two forms
// Form 1: three parameters implied
//  lea eax, [rdi+rdx]

// Form 2: two parameters implied
//   lea eax, [rdi+rsi]

// Test case => sum(5, 6, 7);
\end{lstlisting}
\subsection*{P3: Constants stored outside the function's code}

Apart from immediate values (and constructs like ARM32 jump tables or embedded literal pools), instructions in the \texttt{.text} section use references to access non-immediate data. The actual data typically resides in separate sections such as \texttt{.rodata}, \texttt{.data}, and \texttt{.bss}. In practice, a complete binary includes these sections, but ExeBench's \texttt{asm} field provides only the isolated code of the function. Consequently, these data values cannot be recovered from the dataset's assembly (as shown in \Cref{lst:float_value}).

%Apart from immediate values (and TBB/TBH jump tables in ARM32), the \texttt{.text} section contains only references to data; the actual data resides in \texttt{.rodata}, \texttt{.data}, \texttt{.bss}, etc. This separation is standard. \hlt{In practice, a binary includes these sections, but} ExeBench's \texttt{asm} field provide only the function's code, so these values cannot be recovered from \hlt{the dataset's assembly} (as shown in \Cref{lst:float_value}).

\begin{lstlisting}[language=C, caption={Function with a \texttt{float} value}, label={lst:float_value}]
float sum(int a, int b){
    float c = 74454.235;
    return a+b+c;
}
\end{lstlisting}

For example, line two of \Cref{lst:float_value} assigns the value $74454.235$ to the \lstinline|float| variable \texttt{c}. Once compiled, the value is stored in the \texttt{.rodata} section, and the assembly only references its address. In \Cref{lst:asm_of_float}, the \lstinline|float| value \lstinline|74454.235| is stored as \texttt{1e6b9147} (little-endian) at address \texttt{0x2004}.

\begin{lstlisting}[language={[x64]Assembler}, caption={Assembly code of \Cref{lst:float_value}}, label={lst:asm_of_float}]
<sum>:
endbr64
add    %esi,%edi
pxor   %xmm0,%xmm0
cvtsi2ss %edi,%xmm0
addss  0xea5(%rip),%xmm0        # 2004 <_IO_stdin_used+0x4>
ret

Contents of section .rodata:
2000 01000200 1e6b9147 25660000 00000000
2010 000000c0 132ef240     
\end{lstlisting}

\subsection*{P4: Non-void function without a return}
In ExeBench, about 114k functions are declared non-void (e.g., \lstinline|int, char, float, double,| or their modified forms), but their bodies contain no return statement. Often a parameter is an array and the function operates on it in place. The code compiles, but if a caller uses the return value, the behavior is undefined~\citep{iso_c23}, since no value is returned.

\begin{lstlisting}[language=C, caption={Non-void function without a return statement}, label={lst:non_void_no_ret}]
int sum(int *line, int len){
    int n=0;
    while(n<len){
        line[n]+=5;
        n++;
    }
}
\end{lstlisting}

\Cref{lst:non_void_no_ret} shows one such example: the return type is \textit{int}, but the function contains no return statement.

\subsection*{P5: Dummy placeholder type in real split}
ExeBench is built around stand-alone, executable source-assembly pairs. To achieve this, it mines dependencies from each function's GitHub repository and, when none are found, injects dummy dependencies generated with Psyche-C. Around 9.3\% of the corpus contains functions with dummy user-defined type (UDT) definitions (\Cref{lst:dummy_udt}). Since these placeholder types carry no real content, the source-assembly pair is not faithful to the original program. A test on such a function checks the placeholder rather than the real type.

\begin{lstlisting}[language=C, caption={Dummy UDT dependency}, label={lst:dummy_udt}]
struct rectangle { int dummy; };
struct square { int dummy; };

int sum (struct rectangle a, struct square b){
    ...
}
\end{lstlisting}

\subsection*{P6: Static/inline functions eliminated in optimized assembly}
Some ExeBench functions carry modifiers such as \texttt{static} or \texttt{inline}. These modifiers are common in real-world code. At \texttt{O1} and above, however, the compiler may inline such functions or discard them entirely. No assembly then remains for the target function, so it cannot be evaluated at that level. In ExeBench, 36.3\% of functions are \texttt{static} or \texttt{inline}, and 62\% of these have no code for the target at \texttt{-O3}. 

\begin{lstlisting}[language=C, caption={Static function}, label={lst:static_func}]
static int sum(int a, int b){
    return a+b+5;
}
\end{lstlisting}

For example, \Cref{lst:static_func} defines a function with the \texttt{static} modifier; if the function is trivial, the compiler will constant-folds and inlines it. Without the \texttt{static} modifier, its code is retained, because the compiler must assume it may be called from another \texttt{.c} file via \texttt{extern}.

\subsection*{P7: Functions named \texttt{main}}
Since ExeBench is sourced from real-world GitHub repositories, some functions (6.3\% of the corpus) are named \emph{main}. Extracting assembly from an executable requires linking the function with a driver \emph{main}; when the function itself is named \emph{main}, linking fails with a redefinition error. 

\subsection*{P8: Trivial function bodies}
Some functions have trivial bodies: even though they have parameters and a return type, the body is empty, a single return statement, or at most two lines of code. Although these bodies are valid, around one-third of the functions in \texttt{train\_real\_compilable} are trivial. They are easy to decompile, so a high share of them raises average scores without reflecting harder code.

\subsection*{P9: Truncated definition}
About 41k ExeBench functions (39k of them in \texttt{train\_not\_compilable}) have a truncated definition, cut off either at the start or at the end, so they fail to compile. While checking for truncation, we also find functions written in the legacy K\&R style, where the parameter types are declared after the function signature but before the opening brace. Since C compilers still accept this style, we do not treat it as an issue. 

\begin{lstlisting}[language=json, caption={Truncated function}, label={lst:truncated_func}]
{
    ...
    "func_def":"nt sum(char a, char b){return a+b;}",
    "signature":["int", "char", "char"],
    ...
}
\end{lstlisting}

In ExeBench, the \texttt{signature} field lists the function's return type (at index zero) followed by its parameter types. Some records have truncated types in \texttt{signature}, and some lack a function header entirely, containing only a series of statements without a header or closing brace. 

\section{The \name Dataset}

\label{sec:exebenchplus_dataset}

Based on our analysis of ExeBench, we group the identified issues into two categories. (1) \hlt{Issues where the assembly in ExeBench lacks information that a real-world binary provides, which can affect downstream uses such as LLM-based decompilation} (P1--P3). (2)~Issues that affect what a test outcome means or whether a function can be evaluated at all (P4--P9). \name addresses each with a new field or flag, described below. \hlt{\name preserves ExeBench's licensing and our added fields are released under the MIT license.}

\subsection{Design Principles}
\label{sec:design-principles}

\name follows two design principles. The first is \textbf{additivity}: existing ExeBench fields are not modified; every fix or flag is added as a new field, so \name is a drop-in replacement for ExeBench and existing pipelines keep working. The second is \textbf{flagging over filtering}: for some problems, we add flags instead of modifying or filtering the dataset, leaving downstream users free to decide how to handle each issue. 

\subsection{Schema and New Fields}
\label{sec:schema}

\Cref{tab:schema} lists each new field, its type and the issue it addresses. The \texttt{func\_call} and \texttt{data\_section} fields have a subfield for each architecture (x86, x86-64, ARM32 and ARM64), and \texttt{objdump\_asm} additionally has one per optimization level; the remaining fields are scalar.

\begin{table}
  \centering
  \small
  \begin{tabular}{lll}
    \hline
    \textbf{Field} & \textbf{Type} & \textbf{Issue} \\
    \hline
    \texttt{func\_call} & dict & P1 \\
    \texttt{unused\_param} & bool & P2 \\
    \makecell{\texttt{opt\_real\_exe\_wrapper} \\ \texttt{opt\_angha\_exe\_wrapper}} & str & P2 \\
    \texttt{data\_section} & dict & P3 \\
    \texttt{no\_return} & bool & P4 \\
    \texttt{dummy\_udt} & bool & P5 \\
    \texttt{objdump\_asm} & dict & P6, P7 \\
    \texttt{trivial\_function} & bool & P8 \\
    \texttt{truncated\_funcs} & bool & P9 \\
    \texttt{*\_errors} & str & \\
    \hline
  \end{tabular}
  \caption{New fields introduced by \name.}
  \label{tab:schema}
\end{table}

\subsection{Construction Pipeline}
\label{sec:pipeline}

\begin{figure*}[!t]
  \centering
  \includegraphics[width=\textwidth]{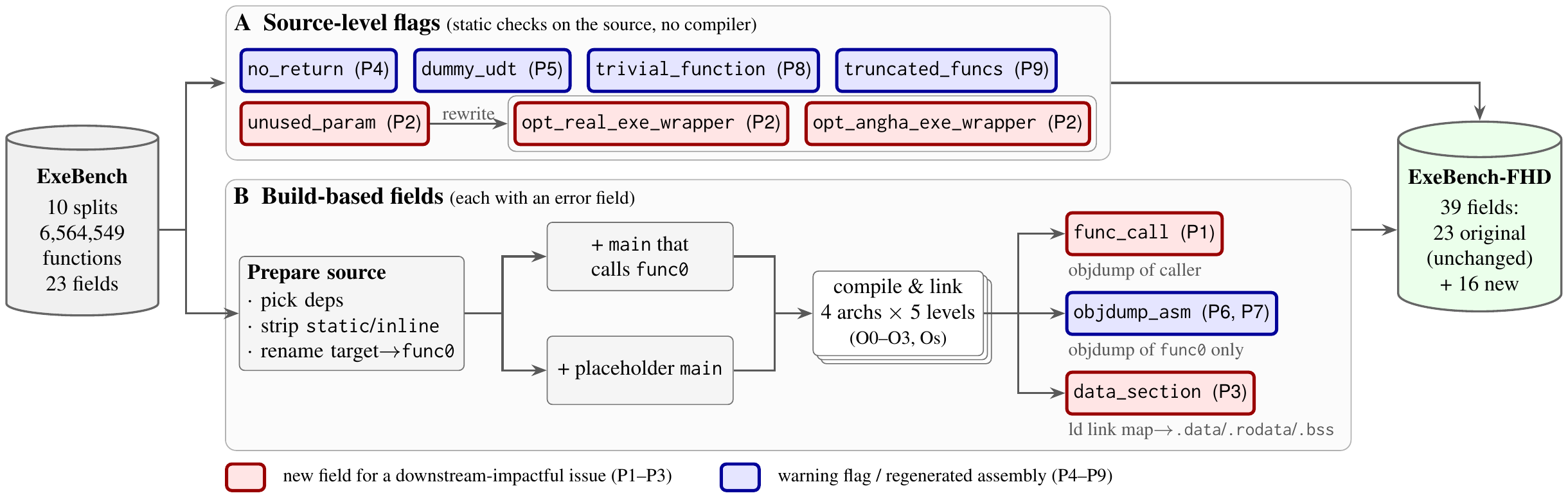}
  \caption{Overview of the \name construction pipeline.}
  \label{fig:pipeline}
\end{figure*}

\Cref{fig:pipeline} gives an overview of the construction pipeline. Each field in \Cref{tab:schema} is implemented as an independent, unit-tested module. Fields such as \texttt{func\_call}, \texttt{data\_section} and \texttt{objdump\_asm} require compiling each function, which makes them the most expensive step in the pipeline. Each of these fields has a companion error field that records what went wrong when the field is empty or null. Once all modules are implemented and tested, a unified script combines them and generates the new splits in a single pass. For \textbf{P6} and \textbf{P7}, the disassembly script strips \texttt{static}/\texttt{inline} modifiers before compiling and omits the driver \emph{main} when the function itself is named \emph{main}.

\subsection{Fixes for Issues That Directly Affect Evaluation}
\label{sec:fixes-impact}

In \name, we first address the issues that directly affect evaluation: P1--P3. 

The main issue is P1: parameter order is ambiguous when a function has at least one floating-point parameter (\lstinline|float, double|) and at least one integer or pointer parameter. To solve this, we introduce a new field, \texttt{func\_call}, which holds the assembly of a call to the function on all four architectures: x86, x86-64, ARM32 and ARM64. On x86, arguments are pushed onto the stack from right to left, so the first parameter ends up at the lowest address; the push order and the callee's stack offsets are therefore enough to recover the parameter order. On the other three architectures, the driver derives each argument from \texttt{argv} in declaration order, so the call site shows which argument is loaded into which register and thus the order of the parameters. 

About 7.6\% of ExeBench functions have unused parameters (P2). In ExeBench, each function has a C++ test wrapper (\texttt{real\_exe\_wrapper}, \texttt{angha\_exe\_wrapper}) that reads the function's IO examples according to its parameter count (from the \texttt{signature} field). When functions are compiled at \texttt{O1} or above, however, the assembly may imply all parameters or only the used ones, depending on the build. A decompiler that recovers only the used parameters fails against this harness, which passes every parameter, used or unused. \hlt{We therefore introduce two new C++ wrappers, \texttt{opt\_real\_exe\_wrapper} and \texttt{opt\_angha\_exe\_wrapper}, that accept either form.

A decompilation passes if it re-executes correctly when called with all parameters or with only the used parameters.} 

As described in P3, data is stored in several sections, and the \texttt{.text} section only contains references to it. We therefore introduce a \texttt{data\_section} field for all four architectures (x86, x86-64, ARM32 and ARM64), which holds the contents of every relevant section that is present (\texttt{.data}, \texttt{.rodata}, \texttt{.bss}, etc.). Since not every record compiles, \texttt{data\_section} is present only for records that build.

\subsection{Fixes for Issues Flagged as Warning}
\label{sec:fixes-warning}

Issues P4--P9 do not hide information from the decompiler, but they affect what a test outcome means or whether a function can be evaluated. For P4, P5, P8 and P9, we add separate flag fields rather than modifying the dataset. These flags let users exclude such functions or report them separately during evaluation. For P6 and P7, a new script (released with the dataset) produces the \texttt{objdump\_asm} field. It regenerates each function's assembly from the linked binary (the \texttt{.out} file), whereas ExeBench's \texttt{asm} field is generated from the \texttt{.o} file. The script supports all four architectures (x86, x86-64, ARM32 and ARM64).

\section{Analysis of \name}
\label{sec:dataset_analysis}

\paragraph{Mixed parameters.} Functions with mixed parameter types (P1) make up 11.5\% of \texttt{test\_real} and 11.1\% of \texttt{valid\_real}, but only 1.4--3.3\% of the other splits listed (\Cref{tab:recoverability} in \Cref{sec:app-mixed}). Some parameter types are hard to build from \texttt{argv}: structures passed by value, SIMD (single instruction, multiple data) vector types and complex numbers. Functions with these types make up 10\% of the corpus but only 0.14\% of \texttt{test\_real}, and \texttt{func\_call} skips them.

\begin{table}[t]
  \centering
  \small
  \resizebox{\columnwidth}{!}{%
  \begin{tabular}{lrrrr}
    \hline
    \textbf{Split} & \textbf{x86} &  \textbf{x86-64} & \textbf{ARM32} & \textbf{ARM64}  \\
    \hline
    test\_real & 9.4 & 15.8 & 2.5 & 5.9 \\
    test\_synth & 27.6 & 28.3 & 26.9 & 27.6 \\
    valid\_real & 10.2 & 16.5 & 3.0 & 6.5 \\
    valid\_synth & 25.9 & 26.5 & 25.2 & 25.8 \\
    train\_real\_compilable & 16.9 & 22.2 & 18.9 & 20.0 \\
    train\_synth\_compilable & 16.4 & 17.4 & 16.6 & 17.0 \\
    \hline
  \end{tabular}}
  \caption{Records with a non-empty data section at \texttt{-O2} (\%). All ten splits in \Cref{tab:app-data-section-full}.}
  \label{tab:data-section}
\end{table}

\paragraph{Data sections.} \Cref{tab:data-section} reports the share of functions in each split whose program has a non-empty data section (P3). On the \texttt{real} splits, the share is lower on ARM32 because ARM32 keeps most constants in small pools inside \texttt{.text} (\emph{literal pools}), where they are already visible in the assembly.

\paragraph{Warning flags.}\label{sec:funnel} Per-split flag rates are in \Cref{tab:flag-rates}. Truncated functions are rare, but the flag helps explain compilation failures. Non-void functions without a return are also rare, but they compile, so without the flag they would be considered in evaluation. A placeholder definition often prevents linking, so \texttt{dummy\_udt} records are less common among disassembled functions (9.3\% of the corpus vs.\ 4.5\% of disassembled records, \Cref{sec:app-flags-objdump}). Removing records with any of the four warning flags keeps 63.9\% of the corpus and 72.2\% of \texttt{test\_real} (\Cref{tab:funnel}). Whether to exclude trivial functions is left to the user. 

The fifth flag, \texttt{unused\_param}, is not part of this filter. Our modified wrapper makes sure that the functions are called correctly during evaluation. One can use this flag only to measure the evaluation efficiency separately (\Cref{sec:downstream-main}).

\section{Impact on Decompiler Evaluation}
\label{sec:downstream-main}

\begin{table}[t]
  \centering
  \small
  \resizebox{\columnwidth}{!}{%
  \begin{tabular}{lrrrr}
    \hline
    \textbf{Subset} & \textbf{x86-64} & \textbf{x86} & \textbf{ARM64} & \textbf{ARM32} \\
    \hline
    Non-empty data sec.\ (P3) & +19.5 & +18.1 & +26.5 & +10.0 \\
    Mixed param.\ types (P1) & +17.0 & +9.8  & +19.1 & +12.5 \\
    Unused param.\ (P2)& +17.6 & +18.0 & +17.1 & +21.0 \\
    \hline
  \end{tabular}}
  \caption{Gain in re-execution rate (percentage) for GPT~6~Sol when \texttt{func\_call} and \texttt{data\_section} are added to the assembly (\texttt{test\_real}, \texttt{-O2}).}
  \label{tab:downstream-main}
\end{table}

\paragraph{Setup.} 

\hlt{
%Recent work fine-tunes open-source models on assembly-source pairs~\citep{nova,llm4decompile,refdecompile}. We instead 
 We use GPT~6~Sol, a frontier commercial model, as such with no training on ExeBench's format for decompilation. If a model of this strength still fails on P1--P3, the cause is the information in the input, not model capacity. We call the model twice for decompilation: once with just the function's assembly (\texttt{asm}) code, and once with the assembly code along with the  \texttt{data\_section} and \texttt{func\_call} (\texttt{+ctx}, where ctx is short for the added context, i.e.\ \texttt{data\_section} and \texttt{func\_call}). The only difference between the two runs is the added context in the input. The results therefore show what the added fields make possible; they do not rank decompilers or show how iterative or agentic pipelines would behave.}

We use \texttt{test\_real}, ExeBench's test split with real dependencies (2,134 functions), compiled with GCC at \texttt{-O2}, a level commonly used for released software. GPT~6~Sol (default settings, one answer per function) decompiles each function twice: under (\texttt{asm}), and (\texttt{+ctx}). We keep every function that can be built and tested on a given architecture, 1,657--1,857 per architecture; functions with an unused parameter, identified by the \texttt{unused\_param} flag, are scored separately. A decompiled function \emph{re-executes} if, compiled into ExeBench's test wrapper, it reproduces the expected output on every IO example; the \emph{re-execution rate} is the share of functions that do. To check that a gain is not due to chance, we use an exact paired sign test (also known as the exact McNemar test), which compares only the functions whose outcome changes between the two runs (details in \Cref{sec:downstream}).

\paragraph{Results.} 

\hlt{\Cref{tab:downstream-main} shows the change in re-execution rate when \texttt{data\_section} and \texttt{func\_call} are added to the assembly, on the functions affected by each issue. Adding the context improves re-execution on every architecture and for every issue: by 10.0--26.5\% on functions with a non-empty data section (P3), 9.8--19.1\% on functions with mixed parameter types (P1), and 17.1--21.0\% on functions with an unused parameter (P2). }

\paragraph{P1: parameter order.} \hlt{For mixed-type parameters, the call-site assembly in \texttt{func\_call} shows which argument is loaded into which register, and thus the parameter order. Without it, the asm-only predictions often have the wrong parameter order or count and do not compile against the test wrapper. With context, re-compilation rises by 13--20\% on the three register architectures, and re-execution rises by 12--19\%. x86, which passes arguments on the stack in order, gains least (+9.8\%) (\Cref{tab:a6-func-call}).}

\paragraph{P2: unused parameters.} \hlt{\name's wrapper (\texttt{opt\_real\_exe\_wrapper}) is executed to test if a decompiled code executes correctly. Our modified wrapper contains two separate function call wrappers. One with all the parameters used in the original source code and one with only the parameters used in the function after compiler optimizations. This enables us to cater for various disassembler behaviours based on what it infers from the assembly pattern implications.  Our wrapper raises the re-execution on functions with unused parameters by 17--21\%, from 58--64\% to 76--82\% (\Cref{tab:a6-unused-param}; $p<10^{-3}$ on every architecture). The call-site assembly helps here because it shows how many arguments the caller passes, which the function's own assembly does not.}

\paragraph{P3: data sections.} Where constants live in data sections, appending \hlt{\texttt{data\_section}} raises re-execution by 18--27\% on x86, x86-64 and ARM64, and by 10\% on ARM32 (\Cref{tab:a6-data-section}).

\hlt{Without the data section, a floating-point constant or a lookup table is visible to the model only as an address, so it has to guess the value. ARM32 has fewer affected functions because it keeps most constants in literal pools inside \texttt{.text}. }

\section{Conclusion}
\label{sec:conclusion}
We analyzed all 6.56M functions in ExeBench and found nine fidelity issues: three cases of missing context, where normal compilation removes information that ExeBench's tests still expect, and six warning-level issues. \name added the missing context as new fields and added warning flags for the rest. The added context raised GPT~6~Sol's re-execution rate by up to 26.5\% on the affected functions. \hlt{Because we focus on evaluation and give a single general-purpose LLM the context in a single pass, the gains isolate the effect of the context itself. This sets up two natural next steps: testing whether agents can retrieve this context on their own, and whether training decompilers with \texttt{func\_call} or \texttt{data\_section} improves them.} We hope this work encourages the community to treat evaluation fidelity as a priority in decompilation benchmarks.

\section*{Limitations}
\label{sec:limitations}
\name has three main limitations. First, every compilation-dependent field (\texttt{func\_call}, \texttt{data\_section} and \texttt{objdump\_asm}) is built with GCC only, so the assembly and data sections may differ from what other compilers such as Clang emit. Second, \name inherits limitations of ExeBench that we leave for future work. About 55\% of records fail to link because ExeBench does not capture their dependencies, so these fields are sparse on the \texttt{synth} splits (\Cref{tab:app-compile-link}). The \texttt{func\_call} generator skips unsupported parameter types and functions named \texttt{main} (16.4\% of the corpus: 10.1\% and 6.3\% respectively, \Cref{tab:app-func-call}) and uses a generated driver instead of the function's real callers. We also flag issues without repairing the source, such as \texttt{signature} fields that disagree with the function definition (\Cref{tab:recoverability}). Third, the warning flags are rules of thumb: \texttt{trivial\_function} depends on a line-count cutoff that is a design choice (\Cref{fig:app-triviality-sweep}), and \texttt{no\_return} and \texttt{truncated\_funcs} use syntactic checks that can miss or misfire on some cases.

\section*{Acknowledgments}
\label{sec:acknowledgments}
We thank the SIT x NVIDIA AI Centre (SNAIC) and the NVIDIA AI Technology Centre (NVAITC) for supporting this research.

\section*{Generative AI Usage}
\label{sec:ai_usage}
We use generative AI tools to paraphrase and improve the language of the manuscript, and to add comments and clean up the code. The ideas, logic, analysis, dataset design and experiments are the authors' own, and the authors review and edit all AI-assisted text and code as needed. The authors take full responsibility for the content of the paper.

\section*{Data and Code Availability}
\label{sec:data_avail}
The scripts used to generate the dataset are available at \url{https://github.com/Akchy/ExeBench-FHD}, and the dataset is available at \url{https://huggingface.co/datasets/davish007/ExeBench-FHD}.

\bibliography{custom}

\newpage

\appendix

\section{Full \name Record Schema}
\label{sec:app-schema}

Every record in the released dataset has the same 39 fields: 23 inherited unchanged from ExeBench and 16 added by \name. Following the additivity principle (\Cref{sec:design-principles}), no inherited field is modified. In particular, ExeBench's \texttt{asm} field is shipped as-is, and the regenerated, binary-level disassembly that addresses P6 and P7 lives in the separate \texttt{objdump\_asm} field. \Cref{tab:app-schema-json,tab:app-schema-string,tab:app-schema-bool} list every field.

\begin{table}[ht]
  \centering
  \small
  \resizebox{\columnwidth}{!}{%
  \begin{tabular}{lll}
    \hline
    \textbf{Field} & \textbf{Origin} & \textbf{Issue} \\
    \hline
    \texttt{asm} & ExeBench & --- \\
    \texttt{real\_iospec}, \texttt{angha\_iospec} & ExeBench & --- \\
    \texttt{real\_io\_pairs}, \texttt{angha\_io\_pairs} & ExeBench & --- \\
    \texttt{func\_call} & \textbf{New} & P1 \\
    \texttt{func\_call\_errors} & \textbf{New} & P1 (diag.) \\
    \texttt{data\_section} & \textbf{New} & P3 \\
    \texttt{data\_section\_errors} & \textbf{New} & P3 (diag.) \\
    \texttt{objdump\_asm} & \textbf{New} & P6, P7 \\
    \texttt{objdump\_asm\_errors} & \textbf{New} & P6, P7 (diag.) \\
    \hline
  \end{tabular}}
  \caption{Fields holding nested JavaScript Object Notation (JSON) objects (11). \emph{diag.}: diagnostic field that records why the main field is empty. \texttt{func\_call}, \texttt{data\_section} and their error fields are indexed by architecture (x86, x86-64, ARM32, ARM64); \texttt{objdump\_asm} and its error field are indexed by architecture and then by optimization level (\texttt{O0}--\texttt{O3}, \texttt{Os}). An error entry is \texttt{null} exactly when that build succeeded.}
  \label{tab:app-schema-json}
\end{table}

\begin{table}[ht]
  \centering
  \small
  \resizebox{\columnwidth}{!}{%
  \begin{tabular}{lll}
    \hline
    \textbf{Field} & \textbf{Origin} & \textbf{Issue} \\
    \hline
    \texttt{path}, \texttt{ref}, \texttt{fname}, \texttt{doc} & ExeBench & --- \\
    \texttt{func\_def}, \texttt{func\_head}, \texttt{func\_head\_types} & ExeBench & --- \\
    \texttt{real\_deps}, \texttt{angha\_deps} & ExeBench & --- \\
    \texttt{real\_error}, \texttt{angha\_error} & ExeBench & --- \\
    \texttt{real\_io\_error}, \texttt{angha\_io\_error} & ExeBench & --- \\
    \texttt{real\_exe\_wrapper}, \texttt{angha\_exe\_wrapper} & ExeBench & --- \\
    \texttt{func\_call\_skip\_reason} & \textbf{New} & P1 (diag.) \\
    \texttt{opt\_real\_exe\_wrapper} & \textbf{New} & P2 \\
    \texttt{opt\_angha\_exe\_wrapper} & \textbf{New} & P2 \\
    \texttt{opt\_real\_exe\_wrapper\_error} & \textbf{New} & P2 (diag.) \\
    \texttt{opt\_angha\_exe\_wrapper\_error} & \textbf{New} & P2 (diag.) \\
    \hline
  \end{tabular}}
  \caption{String fields that may be empty (null) (20).}
  \label{tab:app-schema-string}
\end{table}

\begin{table}[ht]
  \centering
  \small
  \begin{tabular}{lll}
    \hline
    \textbf{Field} & \textbf{Origin} & \textbf{Issue} \\
    \hline
    \texttt{real\_io\_pairs\_are\_trivial} & ExeBench & --- \\
    \texttt{angha\_io\_pairs\_are\_trivial} & ExeBench & --- \\
    \texttt{signature} (list of strings) & ExeBench & --- \\
    \texttt{unused\_param} & \textbf{New} & P2 \\
    \texttt{no\_return} & \textbf{New} & P4 \\
    \texttt{dummy\_udt} & \textbf{New} & P5 \\
    \texttt{trivial\_function} & \textbf{New} & P8 \\
    \texttt{truncated\_funcs} & \textbf{New} & P9 \\
    \hline
  \end{tabular}
  \caption{Boolean fields and the single list-valued field (8).}
  \label{tab:app-schema-bool}
\end{table}

\texttt{func\_call\_skip\_reason} records why \texttt{func\_call} is empty. It takes one of three values: \texttt{unsupported\_param\_type:<kind>} and \texttt{fname\_is\_main} (never attempted, and \texttt{func\_call\_errors} is \texttt{null}), or \texttt{all\_architectures\_failed} (attempted on all four targets, and \texttt{func\_call\_errors} holds the per-architecture reasons).

\section{Mixed Parameter Types per Split}
\label{sec:app-mixed}

\Cref{tab:recoverability} gives the share of functions with mixed parameter types (P1) per split. It also reports how often ExeBench's \texttt{signature} field (the list of the function's return and parameter types) does not match the number of parameters in the definition (\texttt{func\_def}).

\begin{table}[ht]
  \centering
  \small
  \begin{tabular}{lrr}
    \hline
    \textbf{Split} & \textbf{Mixed types} & \textbf{Mismatch} \\
    \hline
    test\_real & 11.5 & 3.8 \\
    test\_synth & 3.3 & 0.1  \\
    valid\_real & 11.1 & 4.2  \\
    valid\_synth & 3.2 & 0.2  \\
    train\_real\_compilable & 2.1 & 1.9  \\
    train\_synth\_compilable & 1.4 & 0.9  \\
    train\_not\_compilable & 1.5 & 10.7  \\
    \hline
  \end{tabular}
  \caption{Mixed parameter types (\%). \textbf{Mismatch}: \texttt{signature} lists a different number of types than the definition.}
  \label{tab:recoverability}
\end{table}

\section{Warning Flags: Extended Analysis}
\label{sec:app-flags}

\begin{table*}[t]
  \centering
  \small
  \begin{tabular}{lrrrrr}
    \hline
    \textbf{Split} & \textbf{no\_ret.} & \textbf{dummy} & \textbf{trivial} & \textbf{trunc.} & \textbf{unused} \\
    \hline
    test\_real                & 0.2 & 0.5  & 27.4 & 0.0 & 7.4  \\
    test\_synth               & 0.3 & 6.2  & 31.7 & 0.0 & 5.9  \\
    valid\_real               & 0.2 & 0.4  & 25.4 & 0.0 & 7.3  \\
    valid\_synth              & 0.3 & 6.0  & 33.0 & 0.0 & 6.2  \\
    train\_real\_compilable   & 1.5 & 4.1  & 36.1 & 0.07 & 6.2  \\
    train\_synth\_compilable  & 0.7 & 11.2 & 24.7 & 0.03 & 6.9  \\
    train\_real\_simple\_io   & 0.4 & 0.7  & 64.5 & 0.0 & 10.5 \\
    train\_synth\_simple\_io  & 0.6 & 15.3 & 45.9 & 0.0 & 8.7  \\
    train\_synth\_rich\_io    & 0.3 & 6.0  & 32.4 & 0.0 & 5.9  \\
    train\_not\_compilable    & 9.4 & 0.0  & 16.8 & 5.45 & 12.1 \\
    \hline
    Whole corpus              & 1.7 & 9.3  & 27.8 & 0.63  & 7.6\\
    \hline
  \end{tabular}
  \caption{Percentage of records with each warning flag set, per split.  Columns are \texttt{no\_return}, \texttt{dummy\_udt}, \texttt{trivial\_function}, \texttt{truncated\_funcs}, \texttt{unused\_param}. }
  \label{tab:flag-rates}
\end{table*}

\subsection{Flag Rates Among Disassembled Functions}
\label{sec:app-flags-objdump}

\Cref{tab:flag-rates} reports each warning flag over every record. A user who evaluates on the regenerated assembly, however, only sees records whose function could be compiled, linked and disassembled. We count a record as \emph{disassembled} if \texttt{objdump\_asm} succeeded for at least one of the 20 (architecture, optimization level) combinations: 2,227,780 records, or 33.9\% of the corpus (\Cref{tab:app-objdump-coverage}). \Cref{tab:app-flag-rates-objdump} gives the flag rates over all records and over this subset.

\begin{table*}[t]
  \centering
  \small
  \begin{tabular}{lrrrr}
    \hline
    \textbf{Split} & \textbf{Records} & \textbf{Disassembled} & \textbf{$\geq$1 of 20 (\%)} & \textbf{All 20 (\%)} \\
    \hline
    test\_real & 2,134 & 2,108 & 98.8 & 85.8 \\
    test\_synth & 5,000 & 2,269 & 45.4 & 44.8 \\
    valid\_real & 2,133 & 2,107 & 98.8 & 86.2 \\
    valid\_synth & 5,000 & 2,225 & 44.5 & 43.8 \\
    train\_real\_compilable & 885,074 & 534,111 & 60.3 & 49.4 \\
    train\_synth\_compilable & 4,159,373 & 1,384,908 & 33.3 & 31.4 \\
    train\_real\_simple\_io & 42,769 & 42,397 & 99.1 & 88.3 \\
    train\_synth\_simple\_io & 640,116 & 207,256 & 32.4 & 31.1 \\
    train\_synth\_rich\_io & 107,250 & 48,745 & 45.4 & 45.1 \\
    train\_not\_compilable & 715,700 & 1,654 & 0.2 & 0.1 \\
    \hline
    Whole corpus & 6,564,549 & 2,227,780 & 33.9 & 31.0 \\
    \hline
  \end{tabular}
  \caption{Coverage of \texttt{objdump\_asm} per split. ``$\geq$1 of 20'' is the share of records disassembled for at least one of the 20 (architecture, optimization level) combinations (x86, x86-64, ARM32, ARM64 $\times$ \texttt{-O0}--\texttt{-O3}, \texttt{-Os}); ``All 20'' is the share that succeeded on every combination.}
  \label{tab:app-objdump-coverage}
\end{table*}

\begin{table*}[t]
  \centering
  \small
  \begin{tabular}{lrrrrr}
    \hline
    \textbf{Split} & \textbf{no\_ret.} & \textbf{dummy} & \textbf{trivial} & \textbf{trunc.} & \textbf{unused} \\
    \hline
    test\_real & 0.2 / 0.2 & 0.5 / 0.5 & 27.4 / 27.5 & 0.0 / 0.0 & 7.4 / 7.3 \\
    test\_synth & 0.3 / 0.4 & 6.2 / 3.5 & 31.7 / 46.3 & 0.0 / 0.0 & 5.9 / 6.9 \\
    valid\_real & 0.2 / 0.2 & 0.4 / 0.4 & 25.4 / 25.3 & 0.0 / 0.0 & 7.3 / 7.3 \\
    valid\_synth & 0.3 / 0.2 & 6.0 / 4.0 & 33.0 / 48.4 & 0.0 / 0.0 & 6.2 / 8.2 \\
    train\_real\_compilable & 1.5 / 1.6 & 4.1 / 1.9 & 36.1 / 39.3 & 0.07 / 0.03 & 6.2 / 7.9 \\
    train\_synth\_compilable & 0.7 / 1.0 & 11.2 / 5.1 & 24.7 / 35.6 & 0.03 / 0.02 & 6.9 / 8.0 \\
    train\_real\_simple\_io & 0.4 / 0.4 & 0.7 / 0.7 & 64.5 / 64.5 & 0.0 / 0.0 & 10.5 / 10.4 \\
    train\_synth\_simple\_io & 0.6 / 1.0 & 15.3 / 8.7 & 45.9 / 61.0 & 0.0 / 0.0 & 8.7 / 15.1 \\
    train\_synth\_rich\_io & 0.3 / 0.3 & 6.0 / 3.4 & 32.4 / 47.5 & 0.0 / 0.0 & 5.9 / 7.0 \\
    train\_not\_compilable & 9.4 / 1.5 & 0.0 / 0.0 & 16.8 / 66.0 & 5.45 / 0.24 & 12.1 / 7.1 \\
    \hline
    Whole corpus & 1.7 / 1.1 & 9.3 / 4.5 & 27.8 / 39.7 & 0.63 / 0.02 & 7.6 / 8.6 \\
    \hline
  \end{tabular}
  \caption{Percentage of records with each warning flag set, as \emph{all records / disassembled records}. The second value is over the records counted in the \texttt{Disassembled} column of \Cref{tab:app-objdump-coverage}. A record can carry several flags.}
  \label{tab:app-flag-rates-objdump}
\end{table*}

The disassembled subset is \emph{more} trivial than the corpus (39.7\% vs.\ 27.8\% flagged \texttt{trivial\_function}), because short, self-contained functions are the ones most likely to build. The effect is largest where few records build: on \texttt{test\_synth} and \texttt{valid\_synth}, where only about 45\% of records disassemble, the trivial share rises by about 15\%. Conversely, \texttt{dummy\_udt} roughly halves (9.3\% to 4.5\%), since a placeholder type often prevents the build altogether, and \texttt{truncated\_funcs} almost disappears (0.016\%, 362 records).

\subsection{Flag Co-occurrence and Compounded Filtering}
\label{sec:app-cooccurrence}

\Cref{tab:app-pairwise} reports how often each pair of warning flags fires on the same record. Over all splits, 61.1\% of records carry none of the five flags, ranging from 34.6\% (\texttt{train\_real\_simple\_io}) to 72.4\% (\texttt{valid\_real}) per split. The most common combination is \texttt{trivial\_function} with \texttt{unused\_param} (3.5\% of the corpus). The next is \texttt{dummy\_udt} with \texttt{trivial\_function} (2.4\%), consistent with dummy-type records skewing toward short bodies.

\begin{table}[ht]
  \centering
  \small
  \begin{tabular}{lrr}
    \hline
    \textbf{Flag pair} & \textbf{Records} & \textbf{\%} \\
    \hline
    trivial \& unused\_param & 229,487 & 3.496 \\
    dummy\_udt \& trivial & 160,783 & 2.449 \\
    dummy\_udt \& unused\_param & 112,514 & 1.714 \\
    no\_return \& trivial & 44,305 & 0.675 \\
    no\_return \& unused\_param & 41,105 & 0.626 \\
    trivial \& truncated & 9,044 & 0.138 \\
    truncated \& unused\_param & 4,534 & 0.069 \\
    no\_return \& truncated & 2,854 & 0.043 \\
    no\_return \& dummy\_udt & 2,209 & 0.034 \\
    dummy\_udt \& truncated & 49 & 0.001 \\
    \hline
  \end{tabular}
  \caption{Pairwise flag co-occurrence over all 6,564,549 records. Percentages are of the full corpus, not conditional on either flag.}
  \label{tab:app-pairwise}
\end{table}

For evaluation, the two flags that matter most are \texttt{dummy\_udt} (the function often cannot be built, and when it can, it tests a placeholder type) and \texttt{trivial\_function} (the function gives a decompiler almost nothing to recover). \Cref{tab:app-usable-eval} applies only these two to the evaluation splits. Over all splits, 4,289,066 records (65.34\%) survive the same filter.

\begin{table}[ht]
  \centering
  \small
  \begin{tabular}{lrrr}
    \hline
    \textbf{Split} & \textbf{Raw} & \textbf{Dropped} & \textbf{Kept (\%)} \\
    \hline
    test\_real & 2,134 & 592 & 72.26 \\
    test\_synth & 5,000 & 1,847 & 63.06 \\
    valid\_real & 2,133 & 547 & 74.36 \\
    valid\_synth & 5,000 & 1,902 & 61.96 \\
    \hline
  \end{tabular}
  \caption{Evaluation-split size after dropping records with \texttt{dummy\_udt} or \texttt{trivial\_function} set.}
  \label{tab:app-usable-eval}
\end{table}

\subsection{Sensitivity of the Triviality Cutoff}
\label{sec:app-triviality}

\begin{figure}[ht]
  \centering
  \begin{tikzpicture}
    \begin{axis}[
      width=\linewidth, height=5.5cm,
      xlabel={Line cutoff}, ylabel={Non-trivial records (\%)},
      xtick={1,2,3,5,10}, ymin=0, ymax=100,
      legend style={font=\tiny, at={(0.5,-0.32)}, anchor=north}, legend columns=2,
      legend cell align={left},
    ]
      \addplot coordinates {(1,83.3)(2,72.8)(3,60.0)(5,44.0)(10,24.8)};
      \addlegendentry{test\_real}
      \addplot coordinates {(1,71.4)(2,64.1)(3,57.8)(5,47.1)(10,29.5)};
      \addlegendentry{train\_real\_compilable}
      \addplot coordinates {(1,43.6)(2,36.1)(3,31.0)(5,22.0)(10,9.3)};
      \addlegendentry{train\_real\_simple\_io}
      \addplot coordinates {(1,87.0)(2,83.1)(3,79.9)(5,73.6)(10,59.5)};
      \addlegendentry{train\_not\_compilable}
    \end{axis}
  \end{tikzpicture}
  \caption{Percentage of records \emph{not} classified as trivial when the triviality rule is ``body has at most $k$ non-blank lines'', for $k \in \{1, 2, 3, 5, 10\}$, on four representative splits. Percentages are over records whose body could be located by matching braces: 99.74\% of the corpus, and 98.3\% of \texttt{train\_not\_compilable}, which holds most truncated definitions. At $k=2$ each curve is within 0.6\% of $100\% - $ the \texttt{trivial\_function} rate in \Cref{tab:flag-rates}. The gap is the multi-line single-\texttt{return} bodies described in \Cref{sec:app-triviality}.}
  \label{fig:app-triviality-sweep}
\end{figure}
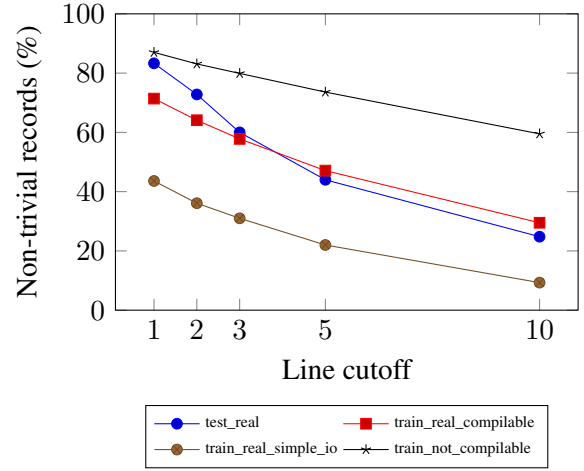

\texttt{trivial\_function} flags a body that is empty, a single \texttt{return} statement, or at most two non-blank, non-comment lines. Over the corpus, every record whose body has at most two lines is flagged (1,803,896 records). The remaining 23,797 flagged records (0.36\%) are single \texttt{return} statements that span more than two lines. \Cref{fig:app-triviality-sweep} shows how the share of \emph{non-trivial} records (those a user keeps after dropping trivial ones) changes as the line cutoff is raised. The share falls steadily, with no point where it suddenly drops, so the cutoff is a trade-off rather than a property of the data. \texttt{train\_real\_simple\_io} is the most trivial split at every cutoff, and \texttt{train\_not\_compilable} the least. Requiring the body to also be branch-free (no \texttt{if}/\texttt{for}/\texttt{while}/\texttt{switch}) flattens the curves: at a 10-line cutoff, 53.4\% of \texttt{test\_real} is still non-trivial, against 24.8\% under the line-count rule without it.

\section{Filtering Funnel and Field Eligibility}
\label{sec:app-funnel}

\begin{table}[ht]
  \centering
  \small
  \begin{tabular}{lrr}
    \hline
    \textbf{Split} & \textbf{Raw} & \textbf{After filter (\%)} \\
    \hline
    test\_real & 2,134 & 72.21 \\
    test\_synth & 5,000 & 62.90 \\
    valid\_real & 2,133 & 74.31  \\
    valid\_synth & 5,000 & 61.72  \\
    train\_real\_compilable & 885,074 & 60.73  \\
    train\_synth\_compilable & 4,159,373 & 66.06 \\
    train\_real\_simple\_io & 42,769 & 35.06 \\
    train\_synth\_simple\_io & 640,116 & 45.50  \\
    train\_synth\_rich\_io & 107,250 & 62.47  \\
    train\_not\_compilable & 715,700 & 73.20  \\
    \hline
    \textbf{Corpus (6,564,549)} & --- & 63.85 \\
    \hline
  \end{tabular}
  \caption{Filtering funnel per split. \texttt{After filter} is the percentage of the split's raw count surviving the drop of all four warning flags (\texttt{dummy\_udt}, \texttt{truncated\_funcs}, \texttt{no\_return}, \texttt{trivial\_function}); \texttt{unused\_param} is not dropped because the either-form wrappers already make those functions evaluable.}
  \label{tab:funnel}
\end{table}

\Cref{tab:app-funnel} extends \Cref{tab:funnel} with the share of each split that can use each compilation-dependent field. The eligibility columns are independent of the filter and of each other. \texttt{func\_call} and \texttt{data\_section} count as eligible when at least one of the four architectures built successfully (for \texttt{data\_section}, this includes builds that simply had no data to report). \texttt{IO-executable} means the record has a non-empty \texttt{real\_io\_pairs} or \texttt{angha\_io\_pairs}.

\begin{table*}[t]
  \centering
  \small
  \resizebox{\textwidth}{!}{%
  \begin{tabular}{lrrrrrr}
    \hline
    \textbf{Split} & \textbf{Raw} & \textbf{$-$dummy\_udt (\%)} & \textbf{All four flags (\%)} & \textbf{func\_call elig.\ (\%)} & \textbf{data\_section elig.\ (\%)} & \textbf{IO-executable (\%)} \\
    \hline
    test\_real & 2,134 & 99.48 & 72.21 & 92.74 & 98.78 & 100.00 \\
    test\_synth & 5,000 & 93.78 & 62.90 & 37.82 & 44.58 & 100.00 \\
    valid\_real & 2,133 & 99.58 & 74.31 & 92.36 & 98.78 & 100.00 \\
    valid\_synth & 5,000 & 93.96 & 61.72 & 36.82 & 43.94 & 100.00 \\
    train\_real\_compilable & 885,074 & 95.90 & 60.73 & 44.29 & 58.47 & 16.11 \\
    train\_synth\_compilable & 4,159,373 & 88.78 & 66.06 & 20.22 & 32.69 & 17.94 \\
    train\_real\_simple\_io & 42,769 & 99.28 & 35.06 & 71.46 & 99.06 & 100.00 \\
    train\_synth\_simple\_io & 640,116 & 84.67 & 45.50 & 21.81 & 31.84 & 100.00 \\
    train\_synth\_rich\_io & 107,250 & 94.05 & 62.47 & 38.22 & 44.68 & 100.00 \\
    train\_not\_compilable & 715,700 & 100.00 & 73.20 & 0.55 & 0.22 & 0.00 \\
    \hline
    \textbf{Corpus} (6,564,549) & --- & 90.73 & 63.85 & 22.17 & 33.23 & 25.79 \\
    \hline
  \end{tabular}}
  \caption{Filtering funnel and per-field eligibility (\emph{elig.}: the field is available for the record). \texttt{$-$dummy\_udt} keeps records without \texttt{dummy\_udt}. \texttt{All four flags} additionally drops \texttt{truncated\_funcs}, \texttt{no\_return} and \texttt{trivial\_function}, and equals the \texttt{After filter} column of \Cref{tab:funnel}. IO examples are present for almost every record, because of how the splits were built, on the evaluation and IO-focused training splits, but not on the two general \texttt{*\_compilable} splits or \texttt{train\_not\_compilable}.}
  \label{tab:app-funnel}
\end{table*}

\section{Effect on Decompiler Evaluation: Details}
\label{sec:downstream}

\paragraph{Setup.}
We decompile the \texttt{-O2} build of every \texttt{test\_real} function that can be prompted and evaluated on a given architecture (it has IO examples, a test wrapper, a successful \texttt{objdump\_asm} and a \texttt{func\_call}), using GPT~6~Sol with its default decoding settings and one sample per function. Each function is decompiled twice: from its disassembly (\emph{asm}), and from the same disassembly followed by its \texttt{data\_section} and \texttt{func\_call} (\emph{+ctx}). The system prompt, the assembly and the output format are identical in both conditions; only the context block differs. A prediction \emph{re-compiles} if it builds inside the function's test wrapper, and \emph{re-executes} if the resulting program reproduces the expected output on every IO example. We use the \texttt{unused\_param} flag to exclude the 144--153 functions per architecture that have an unused parameter from the main results and analyze them separately below. This leaves 1,657 (x86) to 1,857 (x86-64) functions, of which 1,605 are evaluable on all four architectures. Significance is assessed with an exact paired sign test (equivalently, exact McNemar) on the per-function re-execution outcomes.

\paragraph{Overall.}
With context, re-execution improves on every architecture
(\Cref{tab:a6-overall}), by 3.6\% on x86-64, 3.0 on ARM64, 2.8 on x86
and 1.5 on ARM32. The ordering is the same on the 1,605 functions shared by
all four architectures, so it is not caused by the different
numbers of functions per architecture. The gains are significant on x86-64, x86 and ARM64
($p<10^{-3}$) and only just significant on ARM32 ($p=0.049$). Re-compilation barely moves
($\le$2.2\%). The context mainly helps the model produce a function that
behaves correctly, not one that merely compiles. The effect is
concentrated in the functions affected by P1--P3, as the next three paragraphs show.

\paragraph{P3: data sections.}
For the functions that read at least one byte outside \texttt{.text}
(\Cref{tab:a6-data-section}), the context raises re-execution by 19.5\%
on x86-64, 18.1 on x86 and 26.5 on ARM64 ($p<10^{-5}$ each). Without the
data section, a floating-point constant or a lookup table is visible to the
model only as an address, so the model has to guess the value. The ARM32 subset is
small (50 functions), because ARM32 keeps most constants in literal pools
inside \texttt{.text}. Its +10.0\% are not significant ($p=0.27$). The
gain on ARM64 cannot be credited to \texttt{func\_call}, because only 18 of
its 117 data-section functions also have mixed parameters.

\paragraph{P1: parameter order.}
For functions whose signature mixes floating-point and integer (including \texttt{char}) or pointer
parameters (\Cref{tab:a6-func-call}), the context raises re-execution by
19.1\% on ARM64, 17.0 on x86-64 and 12.5 on ARM32 ($p<10^{-3}$ each).
Re-compilation rises by 13--20\% on these three architectures as well. The asm-only predictions
often have the wrong parameter order or number of parameters, so they do not even compile
against the test wrapper. On ARM32 only two of the 199 functions also have a
data section, so the gain there comes from \texttt{func\_call}. On x86-64, by
contrast, 77 of the 199 also have a data section, so part of its gain may be
due to \texttt{data\_section}. x86 is a useful
comparison. Its standard calling convention (cdecl, short for C declaration) passes all arguments on the stack in
declaration order, so the assembly already yields the signature: asm-only
re-compilation is 88.7\% on x86, against 60--63\% on the other three
architectures. Its re-execution gain is correspondingly the smallest
(+9.8\%, $p=0.005$). The remaining gain is likely due to \texttt{func\_call}
also showing the argument types and sizes at the call site.

\paragraph{P2: unused parameters.}
\hlt{\Cref{tab:a6-unused-param} scores the functions with an unused parameter (144--153 per architecture) against the original wrapper, which passes every parameter, and against \name's wrapper, which also accepts a call with only the used parameters. Against the original wrapper alone, the model's re-execution is 28--31\% for asm-only and 63--75\% with context: with the call site it recovers all original parameters, including the unused one. With \name's wrapper, which counts a prediction as correct if it passes either form, re-execution rises from 58--64\% to 76--82\% (17.1--21.0\%, $p<10^{-3}$ on every architecture).}

\begin{table}[ht]
  \centering
  \small
  \resizebox{\columnwidth}{!}{%
  \begin{tabular}{llrrrrrr}
    \hline
    \textbf{Arch} & \textbf{Test form} & & \multicolumn{2}{c}{\textbf{Re-compile (\%)}} & \multicolumn{3}{c}{\textbf{Re-execute (\%)}} \\
     & & $n$ & asm & +ctx & asm & +ctx & $\Delta$ \\
    \hline
    x86-64 & original & 153 & 30.7 & 77.1 & 29.4 & 75.2 & +45.8 \\
           & \name & 153 & 82.4 & 83.0 & 63.4 & 81.0 & +17.6 \\
    \hline
    x86 & original & 144 & 31.2 & 64.6 & 29.2 & 63.2 & +34.0 \\
           & \name & 144 & 84.0 & 84.7 & 63.9 & 81.9 & +18.0 \\
    \hline
    ARM64 & original & 152 & 31.6 & 72.4 & 30.9 & 69.7 & +38.8 \\
           & \name & 152 & 75.0 & 80.9 & 58.6 & 75.7 & +17.1 \\
    \hline
    ARM32 & original & 152 & 29.6 & 73.0 & 27.6 & 71.7 & +44.1 \\
           & \name & 152 & 75.7 & 80.3 & 57.9 & 78.9 & +21.0 \\
    \hline
  \end{tabular}}
  \caption{Functions with a parameter the \texttt{-O2} code never uses, scored against the original wrapper (\texttt{real\_exe\_wrapper}, all parameters) and against \name's wrapper (\texttt{opt\_real\_exe\_wrapper}), which also accepts the used-only call and counts a prediction as correct if it passes either form. With context the model recovers all original parameters from \texttt{func\_call}. The gain is significant on every architecture ($p<10^{-3}$, exact paired sign test).}
  \label{tab:a6-unused-param}
\end{table}

\begin{table}[ht]
  \centering
  \small
  \resizebox{\columnwidth}{!}{%
  \begin{tabular}{lrrrrrr}
    \hline
    \textbf{Arch} & & \multicolumn{2}{c}{\textbf{Re-compile (\%)}} & \multicolumn{3}{c}{\textbf{Re-execute (\%)}} \\
     & $n$ & asm & +ctx & asm & +ctx & $\Delta$ \\
    \hline
    x86-64 & 1857 & 88.0 & 89.8 & 80.5 & 84.1 & +3.6 \\
    x86 & 1657 & 86.9 & 86.8 & 76.0 & 78.8 & +2.8 \\
    ARM64 & 1773 & 86.5 & 87.9 & 73.8 & 76.8 & +3.0 \\
    ARM32 & 1752 & 79.9 & 81.4 & 73.6 & 75.1 & +1.5 \\
    \hline
    \multicolumn{7}{l}{\textit{Functions evaluable on all four architectures}} \\
    x86-64 & 1605 & 89.5 & 91.7 & 81.9 & 86.2 & +4.3 \\
    x86 & 1605 & 86.9 & 86.9 & 76.0 & 78.9 & +2.9 \\
    ARM64 & 1605 & 88.2 & 89.8 & 75.3 & 78.5 & +3.2 \\
    ARM32 & 1605 & 82.1 & 83.6 & 75.9 & 77.3 & +1.4 \\
    \hline
  \end{tabular}}
  \caption{GPT~6~Sol decompilation of \texttt{test\_real} at \texttt{-O2}: assembly only (asm) vs.\ assembly plus \texttt{data\_section} and \texttt{func\_call} (+ctx). Functions with an unused parameter are excluded (see \Cref{tab:a6-unused-param}). Top: every function that can be prompted and evaluated on that architecture; bottom: the 1,605 functions evaluable on all four, so rates are comparable across architectures. $n$: number of functions; $\Delta$: re-execution gain from the context.}
  \label{tab:a6-overall}
\end{table}

\begin{table}[ht]
  \centering
  \small
  \resizebox{\columnwidth}{!}{%
  \begin{tabular}{lrrrrrr}
    \hline
    \textbf{Arch} & & \multicolumn{2}{c}{\textbf{Re-compile (\%)}} & \multicolumn{3}{c}{\textbf{Re-execute (\%)}} \\
     & $n$ & asm & +ctx & asm & +ctx & $\Delta$ \\
    \hline
    x86-64 & 287 & 82.9 & 87.8 & 56.1 & 75.6 & +19.5 \\
    x86 & 177 & 83.6 & 85.3 & 42.9 & 61.0 & +18.1 \\
    ARM64 & 117 & 85.5 & 88.9 & 47.0 & 73.5 & +26.5 \\
    ARM32 & 50 & 72.0 & 74.0 & 46.0 & 56.0 & +10.0 \\
    \hline
  \end{tabular}}
  \caption{Functions that read bytes outside \texttt{.text} (non-empty \texttt{data\_section}), e.g.\ floating-point constants or lookup tables in \texttt{.rodata}. Same setup as \Cref{tab:a6-overall}. ARM32 keeps most constants in literal pools inside \texttt{.text}, and ARM64 builds many constants directly in instructions, hence the smaller $n$.}
  \label{tab:a6-data-section}
\end{table}

\begin{table}[ht]
  \centering
  \small
  \resizebox{\columnwidth}{!}{%
  \begin{tabular}{lrrrrrr}
    \hline
    \textbf{Arch} & & \multicolumn{2}{c}{\textbf{Re-compile (\%)}} & \multicolumn{3}{c}{\textbf{Re-execute (\%)}} \\
     & $n$ & asm & +ctx & asm & +ctx & $\Delta$ \\
    \hline
    x86-64 & 199 & 62.8 & 75.9 & 52.8 & 69.8 & +17.0 \\
    x86 & 195 & 88.7 & 91.3 & 65.6 & 75.4 & +9.8 \\
    ARM64 & 199 & 62.3 & 81.9 & 44.7 & 63.8 & +19.1 \\
    ARM32 & 199 & 59.8 & 73.4 & 51.3 & 63.8 & +12.5 \\
    \hline
  \end{tabular}}
  \caption{Functions whose signature mixes floating-point and integer/pointer parameters, where the function's own assembly does not show the parameter order but the call in \texttt{func\_call} does. 32-bit x86 passes all arguments on the stack in order, so the function's assembly already shows it and the gain is smallest there.}
  \label{tab:a6-func-call}
\end{table}

\subsection{Worked Examples}
\label{sec:app-examples}
\hlt{The two examples below are functions from \texttt{test\_real} where GPT~6~Sol fails from the assembly alone and succeeds with the added context.}

\begin{lstlisting}[float=tb, language={[x64]Assembler}, basicstyle=\ttfamily\scriptsize, numbers=none, aboveskip=4pt, belowskip=0pt, caption={\hlt{P3 example (\texttt{test\_real:2009}, x86-64, \texttt{-O2}). The constant 1.5 is stored in \texttt{.rodata}, so the assembly alone does not show it. GPT~6~Sol's prediction without context compiles but uses the wrong constant.}}, label={lst:ex_data}]
; source: *pdb1 -= 1.5; *pdb2 += 1.5;
movsd  (%rdi),%xmm1
movsd  0xeb8(%rip),%xmm0
subsd  %xmm0,%xmm1
movsd  %xmm1,(%rdi)
addsd  (%rsi),%xmm0
movsd  %xmm0,(%rsi)
ret
; added data_section (.rodata):
;   00 00 00 00 00 00 f8 3f
; asm (fails):  *a -= 1.0; *b += 1.0;
; +ctx (passes): *a -= 1.5; *b += 1.5;
\end{lstlisting}

\begin{lstlisting}[float=tb, language={[x64]Assembler}, basicstyle=\ttfamily\scriptsize, numbers=none, aboveskip=4pt, belowskip=0pt, caption={\hlt{P1 example (\texttt{test\_real:1125}, x86-64, \texttt{-O2}). The \texttt{double} and the pointer use separate register sequences, so the assembly does not show their order. Swapping them makes the test call fail.}}, label={lst:ex_order}]
; source: void f(double x, double *y_pointer)
movapd %xmm0,%xmm1
movsd  0xeb8(%rip),%xmm0
divsd  %xmm1,%xmm0
movsd  %xmm0,(%rdi)
ret
; added func_call:
;   argv[1] -> strtod -> xmm0 (double)
;   argv[2] -> rdi (pointer)
; asm (fails):  void f(double *out, double x)
; +ctx (passes): void f(double x, double *out)
\end{lstlisting}

\section{Build Outcomes and Cross-Architecture Portability}
\label{sec:app-portability}

\subsection{Compile and Link Outcomes}

\Cref{tab:app-compile-link} gives the outcome of building \texttt{data\_section} on each architecture over all records. Every record is attempted, including those named \texttt{main}. Linking, not compiling, is the dominant failure on every architecture. A record fails to link when its function calls another function, or refers to a global, that ExeBench does not capture: this is a consequence of storing one function per record, not of the target.

\begin{table}[ht]
  \centering
  \small
  \resizebox{\columnwidth}{!}{%
  \begin{tabular}{lrrr}
    \hline
    \textbf{Arch} & \textbf{Compile fail (\%)} & \textbf{Link fail (\%)} & \textbf{OK (\%)} \\
    \hline
    x86 & 15.69 & 53.62 & 30.69 \\
    x86-64 & 11.86 & 55.04 & 33.10 \\
    ARM32 & 12.26 & 54.94 & 32.80 \\
    ARM64 & 12.25 & 54.97 & 32.78 \\
    \hline
  \end{tabular}}
  \caption{Build outcome of the \texttt{data\_section} field over all 6,564,549 records, by architecture. Compiler timeouts (60\,s limit) affect fewer than 15 records per architecture and are omitted.}
  \label{tab:app-compile-link}
\end{table}

\Cref{tab:app-data-section-full} extends \Cref{tab:data-section} to all ten splits and adds the build-failure rate. x86 fails more often than x86-64 on every split except \texttt{train\_not\_compilable}, most visibly on the \texttt{real} splits (e.g.\ 9.6\% vs.\ 0.6\% on \texttt{test\_real}). ARM32 and ARM64, by contrast, fail at almost identical rates (within 0.1\%) on every split. \texttt{train\_not\_compilable} almost never builds, as its name says.

\begin{table*}[t]
  \centering
  \small
  \begin{tabular}{lrrrrrrrr}
    \hline
    & \multicolumn{4}{c}{\textbf{Has data (\%)}} & \multicolumn{4}{c}{\textbf{Build fails (\%)}} \\
    \textbf{Split} & \textbf{x86} & \textbf{x86-64} & \textbf{ARM32} & \textbf{ARM64} & \textbf{x86} & \textbf{x86-64} & \textbf{ARM32} & \textbf{ARM64} \\
    \hline
    test\_real & 9.4 & 15.8 & 2.5 & 5.9 & 9.6 & 0.6 & 4.5 & 4.5 \\
    test\_synth & 27.6 & 28.3 & 26.9 & 27.6 & 55.1 & 54.6 & 54.7 & 54.7 \\
    valid\_real & 10.2 & 16.5 & 3.0 & 6.5 & 9.4 & 1.1 & 5.5 & 5.5 \\
    valid\_synth & 25.9 & 26.5 & 25.2 & 25.8 & 55.8 & 55.4 & 55.4 & 55.4 \\
    train\_real\_compilable & 16.9 & 22.2 & 18.9 & 20.0 & 48.9 & 39.8 & 41.4 & 41.4 \\
    train\_synth\_compilable & 16.4 & 17.4 & 16.6 & 17.0 & 68.0 & 66.5 & 66.6 & 66.7 \\
    train\_real\_simple\_io & 8.3 & 10.7 & 5.7 & 7.1 & 10.5 & 0.9 & 2.0 & 2.1 \\
    train\_synth\_simple\_io & 13.9 & 14.8 & 14.2 & 14.4 & 68.8 & 67.6 & 67.6 & 67.6 \\
    train\_synth\_rich\_io & 27.1 & 27.7 & 26.4 & 27.0 & 54.8 & 54.5 & 54.6 & 54.6 \\
    train\_not\_compilable & 0.0 & 0.0 & 0.0 & 0.0 & 99.9 & 99.9 & 99.8 & 99.8 \\
    \hline
  \end{tabular}
  \caption{\texttt{data\_section} outcome per split and architecture at \texttt{-O2}. \texttt{Has data} is the share of records with at least one non-empty data section. \texttt{Build fails} is the share where compiling or linking failed, as opposed to succeeding with no data to report. The remainder built successfully and reference no data outside \texttt{.text}.}
  \label{tab:app-data-section-full}
\end{table*}

The same constant does not always end up as the same bytes in the data section on every architecture. Among records where all four architectures produced a \texttt{.rodata} section, that section is byte-identical on all four in 45.4\% of cases over all splits (208,405 of 459,033 sections). This rises to 63.6\% on \texttt{valid\_real} (28 of 44) and is 42.1\% on \texttt{train\_real\_compilable}. This is expected rather than a defect. ARM can build many constants directly from instruction immediates instead of storing them in memory, and pointer width and alignment differ between the 32-bit and 64-bit targets. This is why \texttt{data\_section} is stored per architecture rather than once per record.

\subsection{Portability of the \texttt{func\_call} Field}

\Cref{tab:app-func-call} breaks the \texttt{func\_call} outcome down by how many of the four architectures succeeded. Over all splits, 7.5\% of records build on all four and 16.4\% are never attempted: 10.1\% have a parameter type the call-site generator does not support (struct or union by value, SIMD vector, \texttt{\_Complex}, etc.) and 6.3\% are functions named \texttt{main}. Records that succeed on exactly two architectures (12.5\% of the corpus) are almost all concentrated in the synthetic-dependency splits. On \texttt{test\_synth} and \texttt{valid\_synth}, every one of the 3,586 two-architecture records succeeds on x86-64 and ARM64 and fails on both 32-bit targets, for a single reason: the synthetic dependency header generated for these splits declares \texttt{typedef unsigned long size\_t}, which conflicts with the 32-bit C library's \texttt{unsigned int} definition. Only 0.15\% of the corpus builds on the x86 targets but not the ARM targets, and 0.01\% on the ARM targets but not the x86 targets. Their errors cluster around inline assembly and special built-in functions for hardware input/output ports that exist on one instruction set only.

\begin{table*}[t]
  \centering
  \small
  \begin{tabular}{lrrrrr}
    \hline
    \textbf{Split} & \textbf{All 4} & \textbf{2--3 of 4} & \textbf{1 of 4} & \textbf{Failed on all} & \textbf{Skipped} \\
    \hline
    test\_real & 80.6 & 10.7 & 1.4 & 7.1 & 0.1 \\
    test\_synth & 1.1 & 36.7 & 0.1 & 53.6 & 8.6 \\
    valid\_real & 80.2 & 10.6 & 1.6 & 7.6 & 0.0 \\
    valid\_synth & 1.0 & 35.8 & 0.0 & 53.3 & 9.9 \\
    train\_real\_compilable & 33.7 & 9.9 & 0.6 & 39.5 & 16.2 \\
    train\_synth\_compilable & 3.4 & 16.8 & 0.1 & 62.9 & 16.8 \\
    train\_real\_simple\_io & 61.2 & 10.0 & 0.2 & 25.9 & 2.6 \\
    train\_synth\_simple\_io & 2.8 & 18.9 & 0.0 & 63.9 & 14.3 \\
    train\_synth\_rich\_io & 1.1 & 37.1 & 0.0 & 52.9 & 8.9 \\
    train\_not\_compilable & 0.4 & 0.1 & 0.1 & 81.7 & 17.7 \\
    \hline
    Whole corpus & 7.5 & 14.6 & 0.15 & 61.5 & 16.4 \\
    \hline
  \end{tabular}
  \caption{Outcome of building \texttt{func\_call} on the four target architectures, as a percentage of each split. \texttt{Skipped} records are never attempted (unsupported parameter type, or a function named \texttt{main}). \texttt{Failed on all} records are attempted on all four architectures and fail on each. Rows sum to 100 up to rounding.}
  \label{tab:app-func-call}
\end{table*}

\section{Size Distribution of \texttt{objdump\_asm}}
\label{sec:app-size-objdump-asm}
Across the 43,348,789 non-empty (record, architecture, optimization level) disassembly instances, the median is 16 instruction lines, the mean 33.7, and the 99th percentile 247. 50,475 records (0.77\%) have at least one instance over 300 lines. \Cref{tab:app-oversized} buckets them by their largest instance. 91.1\% fall between 301 and 1,000 lines and are ordinary large functions, e.g.\ a long formula written out by hand. The largest remaining instance has 254,806 lines (\texttt{Integrand\_3loop\_fun33}, a generated function for numerical integration, at x86-64 \texttt{-O0}).

\subsection{Oversized Assembly Codes}
\label{sec:oversized-asm}

One record, GCC's \texttt{gcc.dg/long-branch-middle.c}, is excluded from \Cref{tab:app-oversized}. Its body is a single inline-assembly directive, \texttt{.fill 32*1024*1024, 4, 0x12345678}, which emits 33,554,432 four-byte words (about 134\,MB) of filler. \texttt{objdump} cannot tell this filler from code and disassembles it into 33.5 million instruction lines, about $10^6$ times the corpus mean of 33.7. We therefore omit \texttt{objdump\_asm} for this record. A search for similar constructs among the 50,475 oversized records (\texttt{.fill}/\texttt{.space}/\texttt{.skip}/\texttt{.zero}/\texttt{.rept} directives, long runs of data directives, or a path inside a compiler test suite) matches only 668 (1.3\%). Of these, 620 are compiler test-suite files and 48 contain data-emitting assembly. The very large functions are therefore mostly genuine code, not artificial compiler stress tests.

\begin{table}[ht]
  \centering
  \small
  \begin{tabular}{lr}
    \hline
    \textbf{Largest instance (lines)} & \textbf{Records} \\
    \hline
    301--1,000 & 45,969 \\
    1,001--5,000 & 4,135 \\
    5,001--10,000 & 220 \\
    10,001--50,000 & 113 \\
    50,001--100,000 & 28 \\
    100,001--500,000 & 10 \\
    \hline
    Total ($>$300) & 50,475 \\
    \hline
  \end{tabular}
  \caption{Records with at least one \texttt{objdump\_asm} instance over 300 instruction lines, bucketed by the record's largest instance over all architectures and optimization levels. The one unusual record whose \texttt{objdump\_asm} is omitted (\Cref{sec:oversized-asm}) is not counted.}
  \label{tab:app-oversized}
\end{table}

\end{document}